\documentclass[12pt,a4paper]{article}

\usepackage{braket}
\usepackage{amsfonts}
\usepackage{color}
\usepackage{authblk}

\newcommand{\beq}{\begin{eqnarray}}
\newcommand{\eeq}{\end{eqnarray}}
\newcommand{\nn}{\nonumber}

\title{Polymeric quantum mechanics and the zeros of the Riemann zeta function}

\author[1]{Jasel Berra-Montiel\thanks{jasel.berra@uaslp.mx}}

\author[1,2]{Alberto Molgado\thanks{alberto.molgado@uaslp.mx}}

\affil[1]{
Facultad de Ciencias, Universidad Autonoma de San Luis Potosi,
Av. Salvador Nava S/N Zona Universitaria,
San Luis Potosi, SLP, 78290 Mexico
}
\affil[2]{
Dual CP Institute of High Energy Physics, Mexico  
}

\begin{document}

\markboth{J.~Berra-Montiel and A.~Molgado}{Polymeric quantum mechanics and the zeros of the Riemann zeta function}

\maketitle

\begin{abstract}
We analyze  the Berry-Keating model and the Sierra and 
Rodr\'iguez-Laguna Hamiltonian within the  polymeric quantization formalism. By using the polymer representation, we obtain for both models, the associated polymeric quantum Hamiltonians and the corresponding stationary wave functions. The self-adjointness condition provide a proper domain for the Hamiltonian operator and the energy spectrum, which turned out to be dependent on an introduced scale parameter. By performing a counting of semiclassical states, we prove that the polymer representation reproduces the smooth part of the Riemann-von Mangoldt formula, and also introduces a correction depending on the energy and the scale parameter.  This may 
shed some light on  the understanding of the fluctuation behavior of the zeros of the 
Riemann function from a purely quantum point of view. 
\end{abstract}





\section{Introduction}
\label{sec:intro}

The Riemann hypothesis is regarded as one of the most important problems in Number Theory \cite{Bombieri}, \cite{Titchmarsh}. It states that the complex non trivial zeros of the classical zeta function all have real part equal to 1/2. First suggested by Riemann himself in his memoirs \cite{Riemann}, the importance of the hypothesis lies on its intrinsic connection with the distribution of the prime numbers, arithmetic functions, quantum information and quantum chaos \cite{Conrey}, \cite{Rosu}. At the beginning of the twentieth century Hilbert and Polya proposed one of the most fascinating approaches to prove the hypothesis, which consists of finding a self-adjoint operator
acting on an appropriate Hilbert space and
whose spectrum agrees with the Riemann zeros \cite{Edwards}, \cite{Elizalde}. Among the reasons behind the Hilbert-Polya conjecture stand out, on the one side, the Selberg trace formula which determines a link between the eigenvalues of the Laplacian operator on Riemannian surfaces and the spectrum of the length associated to their geodesics \cite{Selberg} and, on the other side, the Montgomery-Odlyzko law, which establish that the Riemann zeros are distributed randomly and in accordance with the local statistics of Random Matrix Theory,
which in turn is used to describe the chaotic behaviour of the spectra of atomic nuclei \cite{Mehta}, \cite{Odlyzko}. 
In the same line of thought, Connes proposed to prove the hypothesis using p-adic numbers and adeles.  These structures correspond to an extension of the field of rationals such that the topology is given by a discrete Cantor-like set \cite{Connes}. The p-adic approximation is thus based on the idea that the zeros of the zeta function behave as missing spectral lines of a quantum Hamiltonian given an appropriate semiclassical regularization. Inspired by Connes' approach, Berry put forward the Quantum Chaos conjecture, according to which the Riemann zeros are obtained as the spectrum of a quantum Hamiltonian, whose classical counterpart is given by a chaotic Hamiltonian for which the periodic orbits are completely characterized by the prime numbers \cite{Berry}. Within this framework, Berry and Keating proposed that the classical Hamiltonian $H\!=\!xp$, where $x$ and $p$ correspond to the position and 
momentum for a one dimensional hypothetical particle, respectively, may be intrinsically related to the Riemann zeros \cite{BerryKeat}. This suggestion is based, under certain boundary conditions, on a semiclassical counting of the number of states, which remarkably coincides with the asymptotic behavior of the average number of zeros given by the Riemann-von Mangoldt counting 
formula~\cite{Edwards}. Even though, the model showed to mimic certain interesting features, the Berry-Keating model was unable to derive exactly the Riemann zeros. In an attempt to improve their proposal, Sierra
and Rodr\'iguez-Laguna introduced a generalization of the $xp$-Hamiltonian in such a way that it preserves many of its characteristics, but providing a consistent quantum model which allows for a smooth approximation of the Riemann zeros \cite{Sierra1}.            
Furthermore, these generalizations have allowed to relate the location of the zeros in terms of the massive Dirac equation within the 
context of relativistic spacetimes, conformal quantum mechanics and holographic correspondence \cite{Sierra2}, \cite{Sierra3}, \cite{Sierra4}. Unfortunately, until today there is no known Hamiltonian that meets all the requirements of the Hilbert-Polya conjecture, although recently a non-Hermitian 
$\mathcal{PT}$-symmetric Hamiltonian has been demonstrated 
to formally satisfy those conditions~\cite{bender}. For a more detailed discussion we refer the reader to the review \cite{Sierra5} and references therein.

From a different but related perspective, there have been recent advances in the quantization of classical chaotic systems \cite{Dittrich1}, \cite{Dittrich2}, mainly focused on the strong indication that general relativity proves to be   
a weakly non-integrable or even chaotic 
theory~\cite{Hohn}, \cite{Keifer}. In this case, it is demonstrated that performing a standard quantization could result in the absence of semiclassical states or in the lack of quantum solutions, at least in the case of constrained systems \cite{Dittrich2}. This aspect may be addressed by adapting the topology to the quantization problem, namely, by preserving the continuity properties of the appropriate
classical observables and stating precise rules to promote 
these observables into quantum operators. One way to accomplish this is to use the so-called polymer quantum mechanics, a non-regular, background independent canonical quantization method adapted from Loop Quantum Gravity techniques, which turns out to be not unitarily equivalent to the Schr\"{o}dinger representation \cite{Ashtekar},\cite{Corichi}. 
This quantization scheme introduces a natural scale parameter that adapts to the topology of a given model. 
This parameter results fundamental within the Loop
Quantum Gravity formalism, however, the quantization 
method is not restricted to the 
gravitational field and, in principle, it may be 
relevant for any theory with similar topological structures for which the scale parameter may be
interpreted appropriately.
This particular quantum representation also allows to study the semiclassical limit and the quantum states in a consistent manner with the underlying topological and geometrical aspects defined by the model under consideration. Altogether, according to the Quantum Chaos conjecture, the Riemann zeros are related to chaotic Hamiltonian systems,
thus this type of quantization may allow us to study different aspects of the spectrum of the Riemann zeros, in particular, the zeros smoothness property and also the fluctuation term in the Riemann formula. 

The purpose of this paper is to explore the polymeric quantization of different Hamiltonian models whose spectrum contains the asymptotic approximation of the Riemann zeros. In this sense,  our main goal is to show how the polymer quantization scheme may shed some light on the study of the zeros of the Riemann zeta function, in particular, by obtaining the counting formula of semiclassical states and by inducing a correction term, purely associated to the discretization inherent to the polymeric representation.   Further, as the polymer quantization results are non-equivalent to the Schr\"odinger 
representation, the model here presented may also serve as a motivation to  address different problems in mathematics and theoretical physics from this particular 
quantization scheme.

The paper is organized as follows, in Section 2 we give a brief review of polymeric quantum mechanics in order to set the notational conventions used in the subsequent sections. In Section 3 we apply the polymeric representation to the Berry-Keating $xp$-model, and analyze the properties of the spectrum. In Section 4 a bounded generalization, proposed by Sierra and 
Rodr\'iguez-Laguna in~\cite{Sierra1}, is introduced in order to reproduce the asymptotic behaviour of the zeros and the eigenenergy equation of the Riemann zeta function. Finally, we include some concluding remarks in Section 5. 

\noindent 
\section{Polymer quantum mechanics}
\subsection{Basic notions}

In this section we briefly provide some basic notions concerning the polymer representation of finite dimensional mechanical systems as discussed in~\cite{Ashtekar}, \cite{Corichi}. In the standard Schr\"{o}dinger picture of quantum mechanics the Hilbert space associated to a non relativistic particle on the real line, is given by the square integrable functions with the Lebesgue measure, $L^{2}(\mathbb{R},dx)$. In this case, the position and momentum variables are represented as operators which act as multiplication and differentiation, respectively. However, in order to introduce the polymer representation let us define the abstract kets $\ket{\mu}$ labelled by a real number $\mu$. The space $\mathcal{H}_{poly}$, is a non-separable Hilbert space characterized by an uncountable orthonormal basis given by $\ket{\mu}$ with inner product
\begin{equation}\label{inner}
\braket{\nu|\mu}=\delta_{\nu,\mu},
\end{equation} 
\noindent where $\delta_{\nu,\mu}$ denotes the Kronecker delta function (not a Dirac distribution). Immediately, this implies that a generic vector $\ket{\Psi}\in\mathcal{H}_{poly}$ may be written as $\ket{\Psi}=\sum_{i=1}^{\infty}\braket{\mu_{i}|\Psi}\ket{\mu_{i}}$, for some countable collection of basis kets $\ket{\mu_{i}}$. The state $\ket{\Psi}$ corresponds to a cylinder-like state which occurs naturaly in the loop quantum gravity framework \cite{Ashtekar}. 

\noindent The algebra generated by the operators $\hat{U}_{\nu}=e^{i\nu \hat{x}}$, $\hat{V}_{\lambda}=e^{i\lambda \hat{p}}$, satisfying the product rule
$\hat{U}_{\nu_{1}}\hat{U}_{\nu_{2}}=\hat{U}_{\nu_{1}+\nu_{2}}$, and $\hat{V}_{\mu_{1}}\hat{V}_{\mu_{2}}=\hat{V}_{\mu_{1}+\mu_{2}}$, 
and the canonical commutation relations
\begin{equation}
\hat{U}_{\nu}\hat{V}_{\lambda}=e^{-i\nu\lambda}\hat{V}_{\lambda}\hat{U}_{\nu}.
\end{equation}
\noindent act on the basis as
\beq
\hat{U}_{\nu}\ket{\mu} & = & e^{i\nu\mu}\ket{\mu} \,,\nn\\ 
\hat{V}_{\lambda}\ket{\mu} & = & \ket{\mu-\lambda} \,.
\eeq
\noindent These operators correspond to a one-parameter family of unitary operators defined on $\mathcal{H}_{poly}$, where the position acts as multiplication, precisely as in the Schr\"{o}dinger representation, but the momentum operator is not defined on the polymer Hilbert space, the reason of this feature is a consequence of non-continuity properties of the operator  $\hat{V}_{\lambda}$ in the parameter $\lambda$ with respect to the inner product (\ref{inner}), resulting in the non existence of a Hermitian operator associated to the momentum variable \cite{Stone}.
In order to provide a physical realization corresponding to the abstract Hilbert space $\mathcal{H}_{poly}$, let us introduce the momentum representation, for which the states are identified with plane waves \cite{Tecotl}: 
\begin{equation}
\phi_{\mu}(p):=\braket{p|\mu}=\exp{(i\mu p/\hbar)},
\end{equation}
\noindent where the inner product takes the form
\begin{equation}
\braket{\mu|\nu}=(\phi_{\mu},\phi_{\nu})=\lim_{L\rightarrow\infty}\frac{1}{2L}\int_{-L}^{L}dp\,\phi^{*}_{\mu}(p)\phi_{\nu}(p),
\end{equation}
that is, the polymer Hilbert space in the momentum representation results in the space of almost-periodic functions \cite{Rudin}. This inner product turns out to be equivalent to the inner product defined on a circle in the uniform measure given by \cite{Corichi}
\begin{equation}
\braket{\phi(p)|\psi(p)}_{\mu_{0}}=\frac{\mu_{0}}{2\pi\hbar}\int_{-\pi\hbar/\mu_{0}}^{\pi\hbar/\mu_0}dp\,\phi(p)^{*}\psi(p),
\end{equation}
\noindent  where $\mu_{0}$ is a parameter that characterize the separation between points. This formula remains valid as long as we consider states $\ket{\mu_{n}}$, such that $\mu_{n}=n\mu_{0}$, with $n$ an integer number. The states labelled by $\mu_{0}$ result in a separable Hilbert subspace of the polymer Hilbert space $\mathcal{H}_{poly}$.
\noindent Thus, within this representation, the position operator $\hat{x}$ acts as a derivation $\hat{x}=i\partial/\partial p$, 
while the translation operator $\hat{V}_{\lambda}=e^{i\lambda \hat{p}}$ acts as a simple multiplication by $e^{i\lambda p}$. 

\noindent Given that there is no momentum operator defined in $\mathcal{H}_{poly}$, in order to introduce dynamics, one can define a regulated operator $\hat{p}_{\mu_{0}}$, depending on the scale $\mu_{0}$ as:
\begin{equation}\label{ppoly}
\hat{p}_{\mu_{0}}:=\frac{\hbar}{2i\mu_{0}}\left( \hat{V}_{\mu_{0}}-\hat{V}_{\mu_{0}}^{\dagger}\right)=\frac{\hbar}{\mu_{0}}\sin\left( \frac{\mu_{0}\hat{p}}{\hbar}\right). 
\end{equation}
\noindent In the standard Schr\"{o}dinger representation this operator converges to the usual momentum operator in the limit $\mu_{0}\rightarrow 0$ in  the Lebesgue measure $L^{2}(\mathbb{R},dx)$ but, on the contrary, in the polymer representation the $\mu_{0}\rightarrow 0$ limit in $\mathcal{H}_{poly}$ no longer exists, and therefore, the parameter $\mu_{0}$ can be regarded as a natural scale parameter present in the theory. This is precisely the source of the modification in the energy spectrum of a quantum mechanical system, since any classical function depending on the momenta, whenever it is expressed as a quantum operator on the polymer Hilbert space, will be realized as a scale dependent operator~\cite{Hossain}, feature that will be crucial in the following sections. 

\section{Polymeric quantization of the Berry-Keating-Connes model}
\subsection{Semicalssical approach}
As mentioned before, in 1999 Connes, on the one hand, and Berry and Keating, on the other 
hand, proposed a model which, in a semiclassical approximation, resembles the statistical behaviour of the location of the Riemann zeros in the real line \cite{Connes},\cite{BerryKeat}. This model is determined by the classical Hamiltonian
\begin{equation}\label{Hamxp}
H=xp,
\end{equation}
\noindent where $x$ and $p$ are the position and momentum of a one dimensional particle respectively. The classical solution of the system is given by the family of hyperbolas on the phase space
\begin{equation}
x(t)=x_{0}e^{t}, \;\;\;\; p(t)=p_{0}e^{-t}, 
\end{equation} 
\noindent implying that, at the classical level, the trajectories of the particle are not bounded, thus resulting in that even at the semiclassical level the system possesses a continuous spectrum. It should be noted that even though the Hamiltonian system (\ref{Hamxp}) is not strictly chaotic, it contains a simple class of instability, as the point $x=0$, $p=0$ represents an unstable hyperbolic point, thereby, verifying some of the conditions present in the Quantum Chaos conjecture. In order to solve the problem of the continuous spectrum, Berry and Keating imposed the following constraints in the phase space: $|x|>l_{x}$, $|p|>l_{p}$, where the minimal length in the positions $l_{x}$ and momenta $l_{p}$, span a Planck cell with area $l_{x}l_{p}=2\pi\hbar$. Under this conditions the estimate number of semiclassical energy  states between 0 and $E>0$ is given by
\begin{equation}\label{NE}
\mathcal{N}(E)=\frac{A}{2\pi\hbar},
\end{equation}  
\noindent where $A$, represents the phase space area below the contour allowed by the classical trajectories, subject to the above mentioned constraints. The resulting counting of number of states in the Berry-Keating model, in units $\hbar=1$, reads
\begin{eqnarray}\label{NBK}
\mathcal{N}_{BK}(E)&=&\frac{1}{2\pi}\left[ E\int_{l_{p}}^{E/l_{x}}\frac{dp}{p}-l_{x}\left( \frac{E}{l_{x}}-l_{p}\right) \right], \nonumber \\
 &=&\frac{E}{2\pi}\left( \ln\frac{E}{2\pi}-1\right)+1. 
\end{eqnarray}
\noindent The exact expression for the number 
of states, obtained by Riemann himself \cite{Sierra6}, is given by
\begin{eqnarray}
\mathcal{N}_{R}(E)=\braket{\mathcal{N}(E)}+\mathcal{N}_{fl}(E),
\end{eqnarray}
\noindent where $\braket{\mathcal{N}(E)}$ corresponds to the smooth part of the Riemann-von Mangoldt formula, responsible to provide the average locations of the zeros
\begin{equation}
\braket{\mathcal{N}(E)}\sim \frac{E}{2\pi}\left(\ln\frac{E}{2\pi}-1 \right)+\frac{7}{8}+\ldots,  
\end{equation}
\noindent and a fluctuation term depending on the imaginary part of the zeta function, which can be written as
\begin{equation}\label{fluc}
\mathcal{N}_{fl}(E)=-\frac{1}{\pi}\sum_{p}\sum_{n=1}^{\infty}\frac{1}{np^{n/2}}\sin(nE\ln p),
\end{equation}
\noindent where $p$ denotes the prime numbers. As noted by Berry, this expression behaves as the fluctuation term in the spectrum of a one dimensional chaotic system \cite{Sierra6}. As discussed in this last
reference, in order to obtain the correct constant $7/8$ in equation (\ref{NBK}) a Maslov phase is introduced according to the semiclassical approximation.  However, despite the remarkable connection with the Riemann zeros, the derivation is rather heuristic and the attempts to generalize this approach for the fluctuation term seems to be difficult using the semiclassical approach. Thus, a full quantum version of this model is considered \cite{Sierra5}, \cite{Sierra6}, but even there some connections with the Riemann zeros is lost.  In this manner, taking into consideration the relations between polymer quantization and chaotic classical systems, we explore the polymer quantization of the Berry-Keating model.   In particular, our main intention is to explore the plausibility of obtaining a modification of the spectrum
such that it adjust to the fluctuation term for the distribution 
of the Riemann function zeros.

Before analysing the complete polymer quantum theory, we examine the polymer counting of semiclassical states. In accordance with the polymer representation \cite{Ashtekar},
\cite{Corichi}, the polymer version of the classical Berry-Keating
Hamiltonian~(\ref{Hamxp}) takes the form
\begin{equation}
\label{eq:Hpol}
H_{poly}=\frac{1}{\mu_{0}}x\sin(\mu_{0}p), 
\end{equation}             
which in the limit $\mu_{0}\rightarrow 0$ behaves as the non-polymeric Hamiltonian (\ref{Hamxp}) plus correction terms introduced by the 
 fundamental scale $\mu_0$. By performing an analogous counting of semiclassical states as in (\ref{NE}), given by the area below the classical trajectories resulting from the Hamiltonian~(\ref{eq:Hpol}), we obtain 
\begin{equation}
\mathcal{N}_{poly}(E)=\frac{E}{2\pi}\left(\ln\tan\left( \frac{E\mu_{0}}{2l_{x}}\right) -\ln\tan\left( \frac{\mu_{0}l_{p}}{2} \right)-1 \right), 
\end{equation}
and, by performing an asymptotic analysis for the values $E>>1$, we deduce 
\begin{equation}\label{Npoly}
\mathcal{N}_{poly}(E)\approx\frac{E}{2\pi}\left(\ln\frac{E}{2\pi}-1\right)-\frac{E}{2\pi}\frac{l_{p}^{2}\mu_{0}^{2}}{12}-\frac{E}{2\pi}\frac{7}{1440}l_{p}^{4}\mu_{0}^{4} +O(\mu_{0}^{6}).  
\end{equation}
 It can be seen that the polymer representation reproduces the average number of the Riemann zeros, and also introduces a correction term depending on the energy and the scale parameter $\mu_{0}$. Clearly, the standard counting formula is recovered for $\mu_{0}=0$. From the correction term, it seems that there is a natural minimal length which spans the Planck cell with area $l_{x}l_{p}=2\pi$ (in $\hbar=1$ units), given by $l_{x}=\mu_{0}$. This stems from the fact that in the polymer representation one introduces the structure of a lattice in the configuration space, commonly associated with the length scale $\mu_{0}$ (the lattice spacing), resulting in a regulator of the Planck cells. 
In the literature, perturbations of the counting formula of semiclassical states, are usually related to specific boundary conditions on classical states~\cite{BerryKeat}, \cite{Sierra1}. As we will see in the next section, under the polymer representation, the boundary conditions can be directly related to the energy spectrum through the self-adjoint extension of the Hamiltonian operator defined on $\mathcal{H}_{poly}$.

\subsection{The polymer quantum xp-model} 
 In this section we obtain the polymer quantization of the Berry-Keating model. The starting point is the polymer Hamiltonian operator determined by
\begin{equation}\label{BKpoly}
\hat{H}_{poly}=\frac{\hbar}{\mu_{0}} \hat{x}\left(\hat{V}_{\mu_{0}}-\hat{V}_{\mu_{0}}^{\dagger} \right),   
\end{equation} 
 in this case, the limit $\mu_{0}\rightarrow 0$, does not exist in the Hilbert space $\mathcal{H}_{poly}$, therefore we are not able to remove the scale parameter $\mu_{0}$ in order to determine a unique momentum operator. Following \cite{BerryKeat}, the normal ordered operator associated to (\ref{BKpoly}) reads
\begin{equation}\label{BKpolyo}
\hat{H}_{poly}=\frac{\hbar}{2\mu_{0}}\left[ \hat{x}\left(\hat{V}_{\mu_{0}}-\hat{V}_{\mu_{0}}^{\dagger}\right) +\left( \hat{V}_{\mu_{0}}-\hat{V}_{\mu_{0}}^{\dagger} \right)\hat{x}  \right].
\end{equation} 
 For the purposes of this paper, we consider the momentum representation, in this manner, the position and momentum operator act on the polymer Hilbert space $\mathcal{H}_{poly}$ as
\begin{equation}
\hat{p}_{\mu_{0}}=\frac{\hbar}{\mu_{0}}\sin\left( \frac{\mu_{0}{p}}{\hbar}\right), \;\;\;\;\; \hat{x}=i\hbar\frac{d}{dp}.
\end{equation}  
 If $p\in (0,\pi\hbar/\mu_{0})$, the polymeric 
Hamiltonian~(\ref{BKpolyo}) results equivalent to
\begin{equation}\label{Hsym}
\hat{H}_{poly}=i\hbar\sqrt{\frac{\hbar}{\mu_{0}}\sin\left( \frac{\mu_{0}{p}}{\hbar}\right)}\frac{d}{dp}\sqrt{\frac{\hbar}{\mu_{0}}\sin\left( \frac{\mu_{0}{p}}{\hbar}\right)},
\end{equation}
 which correspond to a symmetric operator defined on proper domain $L^{2}(a,b)$, where $(a,b)\subset(0,\pi\hbar/\mu_{0})$. The energy eigenfunctions, $\hat{H}_{poly}\phi_{E}(p)=E\phi_{E}(p)$, with eigenvalues $E$ are given by
\begin{equation}
\phi_{E}(p)=C\, \frac{\cot\left( \frac{\mu_{0}p}{2\hbar}\right)^{iE/\hbar}}{\sqrt{\sin\left(\frac{\mu_{0}p}{\hbar}\right) }} \,,
\end{equation} 
 where $C$ stands for a constant integration factor. Making an expansion in the parameter $\mu_{0}$ , we deduce the following behavior of the momentum wave function
\begin{equation}
\phi_{E}(p)= \frac{C}{p^{1/2+iE/\hbar}}\frac{\Gamma\left(\frac{1}{4}+\frac{iE}{2\hbar} \right) }{\Gamma\left(\frac{1}{4}-\frac{iE}{2\hbar} \right)}+O(\mu_{0}^{2}) \,.
\end{equation}
The first term corresponds to the Fourier transform of the formal eigenfunctions in the quantum $H=xp$ model studied by Berry and Keating \cite{BerryKeat}, \cite{Sierra5}. In order to calculate the energy eigenvalues, some boundary conditions must be added in order to make the symmetric polymer operator $\hat{H}_{poly}$ a self-adjoint operator. By the von Neumann theorem \cite{Neumann}, this operator admits an infinite number of self-adjoint extensions parametrized by an angle $\theta$, provided that the momentum variable $p$ is defined on a finite interval \cite{Sierra7}. To simplify further calculations, let us choose the self-adjoint domain as
\begin{equation}
\mathcal{D}(\hat{H}_{poly})=\left\lbrace \phi\in L^{2}(m_{1},m_{2}), e^{i\theta}\phi(m_{1})=\sqrt{\sin(\mu_{0}m_{2}/\hbar)}\phi(m_{2}) \right\rbrace, 
\end{equation} 
 where $m_{1}=\pi\hbar/2\mu_{0}<m_{2}$. These boundary conditions set the energy eigenvalues of $\hat{H}_{poly}$,
\begin{equation}
E_{n}=\frac{2\pi\hbar}{\ln\cot\left( \frac{\mu_{0}m_{2}}{2\hbar}\right)}\left(n+\frac{\theta}{2\pi} \right), \;\;\; n\in\mathbb{N}. 
\end{equation}
 Therefore, the spectrum is discrete and the space between levels depends on the scale parameter $\mu_{0}$. By expanding in terms of the parameter $\mu_0$ we deduce that the energy spectrum presents the following behavior
\begin{equation}
E_{n}\sim \frac{2\pi\hbar}{\ln\left(2\hbar/m_{2}\right)}\left( n+\frac{\theta}{2\pi}\right)\left(1+\frac{m_{2}^{2}\mu_{0}^{2}}{12\hbar^{2}\ln(2\hbar/m_{2})}+\ldots \right).
\end{equation} 
The first term in this expression agrees with the resulting spectrum of the Berry-Keating model in the Sierra regularization \cite{Sierra5}, and the remaining terms correspond to corrections to the energy eigenvalues. This terms, resulting from the polymer representation, arise by the self-adjointness condition of the quantum Hamiltonian as an operator defined on the Hilbert space $\mathcal{H}_{poly}$. In this manner, the fluctuation part on the number of density states (\ref{Npoly}), presents a purely quantum origin, as opposed to the classical nature of the fluctuation term resulting from specific boundary conditions on the classical trajectories, as mentioned in \cite{BerryKeat}.  

\section{Polymer quantization of the Sierra and Rodr\'iguez-Laguna model}  
In \cite{Sierra1}, Sierra and Rodr\'iguez-Laguna proposed a modification of the $xp$-Hamiltonian in order to obtain bounded classical trajectories while keeping 
several relevant characteristics of the Berry-Keating model. As they proposed, this was accomplished by introducing the classical Hamiltonian 
\begin{equation}\label{HSierra}
H_{S}=x\left(p+\frac{l_{p}^{2}}{p}\right), \;\;\; x\geq l_{x},\;\; p\in\mathbb{R},  
\end{equation} 
 where the second term in the Hamiltonian together with the minimal length constraint $x\geq l_{x}$ allow to have a discrete spectrum since solutions correspond to closed paths in phase space. 
Even though the introduction of a minimal length may originate
a modified uncertainty relation at the quantum 
level~\cite{kempf}, we want to emphasize that the main purpose of the minimal 
length considered in the Hamiltonian~(\ref{HSierra}) is, according to~\cite{BerryKeat},  to regularize the system in 
order to obtain a classical bounded motion with a 
correspondent discrete quantum spectrum.  However, it is important to mention that the polymer representation is consequent with a minimal length scale, and thus it brings in a 
natural way a modified uncertainty relation which may be 
related to the one introduced in~\cite{kempf} in an appropriate 
limit~\cite{hussain}.  Below we will focus on completing 
the polymer 
quantization and studying the density of energy states, leaving aside 
issues related to quantum states with a well defined uncertainty position.
The aim of this section is to analyze the properties of a polymer quantization of the Sierra and 
 Rodr\'iguez-Laguna model in order to obtain the energy spectrum. Let us start with the polymer Hamiltonian version of the Sierra and Rodr\'iguez-Laguna model in the momentum representation. By expressing the Hamiltonian (\ref{HSierra}) in terms of the position and momentum as a regulated operators, defined in the polymeric representation (\ref{ppoly}),  we obtain in $\hbar=1$ units
\begin{equation}\label{HSpoly}
\hat{H}_{S_{poly}}=i\frac{d}{dp}\left[\left( \frac{2-2\cos(\mu_{0} p)}{\mu_{0}^{2}}+l_{p}^{2}\right)\frac{\mu_{0}}{\sin(\mu_{0}p)} \right] \,,
\end{equation}
where the cosine term stands for the quadratic momentum term in the polymer representation.  
As we can observe, this operator does not correspond to a symmetric operator, but, after choosing an analogous ordering prescription to the case of the Hamiltonian (\ref{Hsym}), the operator $\hat{H}_{S_{poly}}$ can be rendered as a symmetric operator, whose energy eigenfunctions are given by
\begin{equation}\label{WFS}
\psi_{E}(p)=N\, \frac{\sin(\mu_{0}p)^{1/2}}{\left( 2+\mu_{0}^{2}l_{p}^{2}-2\cos(\mu_{0}p)\right)^{\frac{1}{2}+\frac{iE}{2}}} \,,
\end{equation}
 where $N$ is a constant integration factor. An expansion in the scale parameter $\mu_{0}$ leads to the following expression
\begin{equation}\label{WFSA}
\psi_{E}(p)=\frac{N}{\left(p^{2}+l_{p}^{2}\right)^{\frac{iE}{2}}}\sqrt{\frac{p}{p^{2}+l_{p}^{2}}}+O(\mu_{0}^{3}). 
\end{equation}
 For large values of $p$, the wave functions (\ref{WFSA}) behave as the wave functions originated
 in the quantization of the $xp$-Hamiltonian as, within this limit, the Sierra and Rodr\'iguez-Laguna Hamiltonian (\ref{HSierra}) tends to the $xp$-Hamiltonian. However, we must note that this expansion remains valid for any value of the momentum $p$.  With the aim of deriving 
the corresponding eigenvalues, we must select
as for the $xp$-model a proper domain in order to obtain a correct self-adjoint operator from the Hamiltonian operator defined in (\ref{HSpoly}).
In this sense, using the von Neumann theorem, let us take such a domain as
   \begin{equation}
\mathcal{D}(\hat{H}_{S_{poly}})=\left\lbrace \psi\in L^{2}(m_{1},m_{2}), e^{i\theta}\psi(m_{1})=\frac{\csc({m_{2}})}{\Delta_{\mu_0}}\psi(m_{2}) \right\rbrace \,, 
\end{equation} 
 where $m_{1}=\pi/2\mu_{0}<m_{2}$ and 
$\Delta_{\mu_0}$ is defined by the 
constant expression
\beq
\Delta_{\mu_{0}} :=\sqrt{\frac{2+\mu_{0}^{2}l_{p}^{2}}{2+\mu_{0}^{2}l_{p}^{2}-2\cos(m_{2})}} \,.
\eeq 
Within this boundary conditions, it can 
be shown that the energy eigenvalues are discrete and take the following values
\begin{equation}
E_{n}=\frac{2\pi}{\ln\Delta_{\mu_{0}}}\left( n+\frac{\theta}{2\pi}\right) \,.  
\end{equation}
 By expanding in the parameter $\mu_{0}$, the energy spectrum presents the following behaviour
\begin{equation}
E_{n}=\frac{4\pi}{\ln\left( \frac{1}{1-cos(m_{2})}\right) }\left(n+\frac{\theta}{2\pi} \right) \left( 1+\frac{l_{p}^{2}\mu_{0}^{2}\cos(m_{2})}{(1-\cos(m_{2}))\ln\left(\frac{1}{1-cos(m_{2})} \right) }+\ldots\right), 
\end{equation}

\noindent Similarly to the $xp$-Hamiltonian (\ref{Npoly}), the polymer representation introduces some corrections to the energy spectrum depending on the scale parameter. This correction terms arise from the definition of the self-adjoint extension of the quantum Hamiltonian $\hat{H}_{S_{poly}}$, and therefore, thus they represent a purely quantum behaviour.
Finally, using the classical trajectories derived by the classical counterpart of the polymer Hamiltonian (\ref{HSpoly}), and performing an analogous counting of semiclassical states as in (\ref{NE}), by performing an asymptotic expansion with $E>>1$ for the density of energy states we obtain the expression 
\begin{equation}
\mathcal{N}_{S_{poly}}(E)\approx \frac{E}{2\pi}\left(\ln\frac{E}{2\pi}-1 \right)-\frac{E}{2\pi}\frac{\mu_{0}^{2}\,l_{p}^{2}}{12}+O(\mu_{0}^{4}).  
\end{equation}

\noindent As in the $xp$-Hamiltonian, this expression reproduces the smooth part of the Riemann-von Mangoldt formula and introduces some corrections depending linearly on the energy $E$ and the scale parameter $\mu_0$, which can be related to a correction term in the average location of the Riemann zeros.

In this way, we conclude that the direct responsibility of the appearance of these correction terms lies on the definition of the polymeric representation, since the configuration space is basically defined on a regular graph, which consists of a set of numerable equidistant points, separated by a constant parameter $\mu_{0}$. This means, that in order to define a regulated operator $\hat{p}_{\mu_0}$, we need to approximate it by using trigonometric functions $e^{i\mu_{0}p/\hbar}$ that depends on the scale parameter $\mu_{0}$. This regularization also introduce some corrections terms into the energy eigenvalue equation modifying the spectrum. In order to obtain a more accurate expression related to the fluctuation part $\mathcal{N}_{fl}(E)$ in equation (\ref{fluc}), is probable that a different discretization must be taken into account, since a set of equidistant points, as introduced in our polymeric 
quantum scheme, may not reflect the appropriate properties of the prime numbers, which are intrinsically related to the zeros of the Riemann function on the configuration space. 
In order to deal with this issue, one may equip the configuration space of the polymer representation with a $p$-adic structure~\cite{Connes}, this means that the uniform discretization is replaced by a Cantor-like discretization based on the prime numbers, giving a spectral interpretation of the zeros in terms of functions defined on adele classes. Another direction  may be to consider the scale $\mu_{0}$, present in the polymer quantum Hamiltonians, as a free parameter that can be tuned using variational techniques, 
for example, in order to improve the approximation of the counting of the density of states \cite{Paolo, kleinert}. Further investigation will be necessary in order to clarify these issues. This will be done elsewhere.

\section{Conclusions}

In this paper we analysed two different Hamiltonian models related to the zeros of the Riemann zeta function within the  polymeric quantization formalism. On the one hand, the Berry-Keating Hamiltonian corresponds to a one dimensional unbounded system whose classical trajectories are given by hyperbolas in phase space, on the other hand, the Sierra and Rodr\'iguez-Laguna Hamiltonian consists in a modified $xp$-model with bounded and periodic orbits. By using the polymer representation, we obtained for both models, the associated polymeric quantum Hamiltonians and the corresponding stationary wave functions. In particular, the self-adjointness condition provided a proper domain for the Hamiltonian operator and the energy spectrum, which turned out to be dependent on the scale parameter $\mu_{0}$. Carrying out a counting of semiclassical states, the polymer representation reproduces the smooth part of the Riemann-von Mangoldt formula, and introduces a correction depending on the energy and the scale parameter which may throw some light on the understanding of 
 the fluctuation behaviour of the Riemann zeros. 
From our perspective, the correction term 
is originated by the quantum representation in $\mathcal{H}_{poly}$, as opposed to fixing certain boundary conditions for the classical trajectories, 
as presented in the literature. We expect the results established here may clarify the physical interpretation of the Riemann zeros and encourage  the research in loop and polymer representation to tackle different problems in mathematics and theoretical physics.

\section*{Acknowledgements}
The authors would like to acknowledge financial support from CONACYT-Mexico under projects CB-2014-243433 and CB-2017-283838.


\end{document}